# PFL: a Probabilistic Logic for Fault Trees*

Stefano M. Nicoletti[1], Milan Lopuhaä-Zwakenberg[1], E. Moritz Hahn[1], and Mariëlle Stoelinga[1,2]

[1] Formal Methods and Tools, University of Twente, Enschede, the Netherlands
{s.m.nicoletti,m.a.lopuhaa,e.m.hahn,m.i.a.stoelinga}@utwente.nl
[2] Department of Software Science, Radboud University, Nijmegen, the Netherlands

**Abstract** Safety-critical infrastructures must operate in a safe and reliable way. Fault tree analysis is a widespread method used for risk assessment of these systems: fault trees (FTs) are required by, e.g., the Federal Aviation Administration and the Nuclear Regulatory Commission. In spite of their popularity, little work has been done on formulating structural queries about FT and analyzing these, e.g., when evaluating potential scenarios, and to give practitioners instruments to formulate queries on FTs in an understandable yet powerful way. In this paper, we aim to fill this gap by extending *BFL* [32], a logic that reasons about Boolean FTs. To do so, we introduce a Probabilistic Fault tree Logic (PFL). PFL is a simple, yet expressive logic that supports easier formulation of complex scenarios and specification of FT properties that comprise probabilities. Alongside PFL, we present LangPFL, a domain specific language to further ease property specification. We showcase PFL and LangPFL by applying them to a COVID-19 related FT and to a FT for an oil/gas pipeline. Finally, we present theory and model checking algorithms based on binary decision diagrams (BDDs).

## 1 Introduction

Our self-driving cars, power plants, oil/gas refineries and transportation systems must operate in a safe and reliable way. Risk assessment is a key activity to identify, analyze and prioritize the risk in a system, and come up with (cost-)effective countermeasures. Fault tree analysis (FTA) [38, 40] is a widespread formalism to support risk assessment. FTA is applied to many safety-critical systems and the use of fault trees is required, e.g., by the Federal Aviation Administration (FAA), the Nuclear Regulatory Commission (NRC), in the

* This work was partially funded by the NWO grant NWA.1160.18.238 (PrimaVera), and the European Union's Horizon 2020 research and innovation programme under the Marie Skłodowska-Curie grant agreement No 101008233, and the ERC Consolidator Grant 864075 (*CAESAR*). This preprint has not undergone peer review (when applicable) or any post-submission improvements or corrections. The Version of Record of this contribution is published in the Proceedings of the 25th International Symposium on Formal Methods (LNCS, volume 14000), and is available online at https://doi.org/10.1007/978-3-031-27481-7_13.

ISO 26262 standard [26] for autonomous driving and for software development in aerospace systems. A fault tree (FT) models how component failures arise and propagate through the system, eventually leading to system level failures. Leaves in a FT represent *basic events* (BEs), i.e. elements of the tree that do not need further refinement. Once these fail, the failure is propagated through the *intermediate events* (IEs) via *gates*, to eventually reach the *top level event* (TLE), which symbolizes system failure. In the (sub)tree represented in Fig. 1, the TLE— *Medium Corrosion* — is refined by an AND-gate ($MeC$). For $MeC$ to fail, water must be present, i.e., the *With Water* ($WW$) BE must fail, and there must be at least one acid medium in the pipes, i.e., *Acid Medium* ($AcM$) has to happen. This last OR-gate is further refined with three BEs: for it to fail, at least one of its three children needs to fail. This means that either *Hydrogen sulfide* ($H_2S$) or *Oxygen* ($O_2$) or *Carbon dioxide* ($CO_2$) must be present. Fault tree analysis supports qualitative and quantitative analysis. Qualitative analysis aims at pointing out root causes and critical paths in the system. One can identify the *minimal cut sets* (MCSs) of a FT, i.e. minimal sets of BEs that, when failed, cause the system to fail. One can also identify *minimal path sets* (MPSs), i.e. minimal sets of BEs that - when operational - guarantee that the system will remain operational. Quantitative analysis allows to compute relevant dependability metrics, such as the system reliability, availability and mean time to failure. A formal background on FTs is given in Sec. 2.

| **Medium Corrosion:** | |
|---|---|
| Medium Corrosion | MeC |
| With Water | WW |
| Acid Medium | AcM |
| $H_2S$ | $H_2S$ |
| $O_2$ | $O_2$ |
| $CO_2$ | $CO_2$ |

Figure 1: FT excerpt from Fig. 3.

**Probabilistic Fault tree Logic.** In spite of their popularity, little work has been done on formulating structural queries about FTs and analyzing these, e.g., when evaluating potential scenarios, and to give practitioners instruments to formulate queries on FTs in an understandable yet powerful way. Usually, FTs are translated to stochastic models and existing logics specify properties on these, rather than on elements of FTs. An exception [32] presents a logic to reason about static FTs when BEs have Boolean values. The present work aims to extend that framework by devising a probabilistic logic for FTs, called PFL, where one could easily reason about FTs also taking probabilities into account. To further meet the need for usability - that we uncovered through interviews with a domain expert [3] - we present a domain specific language for PFL, LangPFL, and showcase property specification with both on two case studies, one with a COVID-19 FT, and one with an oil/gas pipeline FT.

**Model checking.** In this paper, we provide model checking algorithms that extend the work in [32]. While we build from algorithms from [32], we require extensions for formulae in which probabilities come into play. We introduce novel algorithms which can decide 1. whether a single probability assignment to all BEs of a FT satisfy a formula; 2. whether a formula is satisfied for all possible probability assignments to BEs and 3. in which regions of the parameter space the considered formula holds. In continuity with previous work, all three algorithms are based on construction and manipulation of binary decision diagrams (BDDs).

**Related work.** Numerous logics describe properties of state-transition systems, such as labelled transition systems (LTSs) and Markov models, e.g., CTL [14], LTL [35], and their variants for Markov models, PCTL [25] and PLTL [33]. State-transition systems are usually not written by hand, but are the result of the semantics of high-level description mechanisms, such as AADL [9], the hardware description language VHDL [19] or model description languages such as JANI [11] or PRISM [30]. Consequently, these logics are not used to reason about the structure of such models (e.g. the placement of circuit elements in a VHDL model or the structure of modules in a PRISM model), but on the temporal behaviour of the underlying state-transition system. Similarly, related work on model checking on FTs [41, 6, 42, 8] exhibits significant differences: these works perform model checking by referring to states in the underlying stochastic models, and properties are formulated in terms of these stochastic logics, not in terms of events in the given FT. In [43], the author provides a formulation of *Pandora*, a logic for the qualitative analysis of temporal FTs. In spite of the use of logic to capture properties of FT, [43] focuses on the analysis of time, introducing gates that are different from the ones considered in this work: the Priority-AND-gate (PAND), the Simultaneous-AND-gate (SAND), and the Priority-OR gate (POR). In PFL we do not (yet) consider time and we focus on AND, OR and VOT-gates. Furthermore, [43] focuses more on the algorithmic part of FTA while leaving out any formalization of FTs or the logic defined upon. Another exception is [32], presenting *BFL*, a logic on FTs that however reasons about FTs only in Boolean terms. We take this framework and develop a logic that extends *BFL* with probabilities. Literature related to FTs, property specification languages, BDDs and parametric model checking is referenced and contextualized in Sec. 2, Sec. 5, Sec. 6.1 and Sec. 6.5.

**Contributions.** To summarize, in this work:
1. We develop PFL, a probabilistic logic to reason about FTs.
2. We present a domain specific language for PFL, LangPFL, to further ease property specification.
3. We showcase the potential of PFL and LangPFL by applying them to a medium-sized COVID-19 related example and to a large-sized case study of an oil/gas pipeline.
4. We provide model checking algorithms to check properties defined in PFL.
5. We provide the theory and an algorithm to solve problems where the probabilities of BEs are parametric.

**Structure of the paper.** Sec. 2 covers background on FTs, Sec. 3 describes PFL, Sec. 4 shows the application of PFL to case studies, Sec. 5 introduces LangPFL, Sec. 6 presents algorithms and Sec. 7 concludes our work.

## 2 Fault Trees: Background

Developed in the early '60s [21], FTs are directed acyclic graphs (DAGs) that model how low-level failures can propagate and cause a system-level failure. The overall failure of a system is captured by a *top level event* (TLE), that is refined through the use of *gates*. FTs come with different gate types. For the purposes of our paper, we will focus on *static* fault trees, featuring OR-gates, AND-gates and VOT*(k/N)*-gates. For a low-level failure to propagate, at least one child of an OR-gate has to fail, all the children of an AND-gate must fail, and at least $k$ out of $N$ children must fail for a VOT(k/N)-gate to fail. When gates can no longer be refined, we reach the *basic events* (BEs) which are the leaves of the tree. FTs enable both qualitative and quantitative analyses. On the qualitative side, *minimal cut sets* (MCSs) and *minimal path sets* (MPSs) highlight root causes of failures and critical paths in the system. MCSs are minimal sets of events that - when failed - cause the failure of the TLE. MPSs are minimal sets of events that - when remaining operational - guarantee that the TLE will remain operational.

**Definition 1 (*Fault Tree*).** *A* Fault Tree *is a tuple* $T = (\mathtt{E}, A, t)$ *where* $(\mathtt{E}, A)$ *is a rooted directed acyclic graph (*$\mathtt{E}$ *are the vertices, called* events*) and* $t$ *is a map* $\mathtt{E} \to \{\mathtt{AND}, \mathtt{OR}, \mathtt{BE}\}$ *such that* $t(e) = \mathtt{BE}$ *iff* $e$ *is a leaf.*

We denote the top event by $e_{top}$, and the set of children of an event $e$ by $ch(e) = \{e' \mid (e, e') \in A\}$. Slightly abusing notation, we denote the set of *basic events*, $e$ with $t(e) = \mathtt{BE}$, as $\mathtt{BE}$, whose elements we enumerate $\mathtt{BE} = \{e_1, \ldots, e_n\}$. We also define the set of intermediate events $\mathtt{IE} = \mathtt{E} \setminus \mathtt{BE}$. The behaviour of a FT $T$ can be rigorously expressed through its *structure function* [38] - $\Phi_T$: if we assume the convention that a BE has value 1 if failed and 0 if operational, the structure function indicates the status of the TLE given the status of all the $n$ BEs of $T$, given by a Boolean vector $\overline{b} = (b_1, \ldots, b_n)$.

**Definition 2 (*Structure Function*).** *The structure function of an FT* $T$ *is a function* $\Phi_T \colon \mathbb{B}^n \times \mathtt{E} \to \mathbb{B}$ *defined recursively by*

$$\Phi_T(\overline{b}, e) = \begin{cases} b_i & \text{if } e = e_i \in \mathtt{BE} \\ \bigvee_{e' \in ch(e)} \Phi_T(\overline{b}, e') & \text{if } t(e)=\mathtt{OR} \\ \bigwedge_{e' \in ch(e)} \Phi_T(\overline{b}, e') & \text{if } t(e)=\mathtt{AND} \end{cases}$$

Thus, for each set of BEs we can identify its characteristic vector $\overline{b}$. One can extend Def. 2 by allowing voting gates, where a gate with $t(e)$=$\mathtt{VOT}(k/N)$ fails if at least k of its children fail, i.e.

$$\sum_{e' \in ch(e)} \Phi_T(\overline{b}, e') \geq k$$

We can also define the classical notions of minimal cut sets and minimal path sets [38]. A cut set is any set of basic events that causes the TLE to occur, i.e., for which the structure function evaluates to 1. A path set is any set of basic events that does not cause the TLE to occur, i.e., for which the structure function evaluates to 0.

**Definition 3.** *A status vector* $\overline{b}$ *is a* cut set (CS) *for* $e \in \mathtt{E}$ *of a given tree* $T$ *if* $\Phi_T(\overline{b}, e) = 1$. *A* minimal cut set (MCS) *is a cut set of which no subset is a cut set:* $\overline{b}$ *is a MCS for* $e \in \mathtt{E}$ *of* $T$ *if* $\Phi_T(\overline{b}, e) = 1 \wedge \forall \overline{b'} \subset \overline{b}, \Phi_T(\overline{b}, e) = 0$.

**Definition 4.** *A status vector* $\overline{b}$ *is a* path set (PS) *for* $e \in \mathtt{E}$ *of a given tree* $T$ *if* $\Phi_T(\overline{b}, e) = 0$. *A* minimal path set (MPS) *is a path set of which no subset is a path set:* $\overline{b}$ *is a MPS for* $e \in \mathtt{E}$ *of* $T$ *if* $\Phi_T(\overline{b}, e) = 0 \wedge \forall \overline{b'} \subset \overline{b}, \Phi_T(\overline{b}, e) = 1$.

# 3 A Probabilistic Logic to Reason about FTs

## 3.1 Syntax

Our logic - PFL - consists of three syntactical layers that we represent with $\phi$, $\psi$ and $\xi$ respectively. To refer to layer two or layer three formulae indistinctly we simply write $\theta$. $\chi$ is a generic formula in PFL. We indicate atomic formulae with the letter $e$: each atomic formula represent an element of a given FT, it being an IE or a BE. Formulae in $\phi$ and $\psi$ can be rewritten with the usual negation and conjunction. Furthermore, in layer one we have the possibility to arbitrarily set the value of one atom $e$ in complex formulae either to 0 or to 1 by writing $\phi[e \mapsto 0]$ and $\phi[e \mapsto 1]$. Please note that $\phi[e \mapsto 0]$ is not equivalent to $\phi \wedge \neg e$. This can be clearly shown if we assume $\phi = \neg e$. In this scenario, we would have $(\neg e)[e \mapsto 0] = \mathit{true}$ while $(\neg e) \wedge \neg e$ would not necessarily equal *true*. Moreover, we have operators to check for MPSs and MCSs for a given formula. The second layer allows us to reason about probabilities. We can check whether the probability of a given layer one formula (potentially conditioned by another one) respects a certain threshold. We have the possibility to set the value of one atom $e$ in complex formulae to an arbitrary probability value $p$. We can also check if two formulae (e.g., two intermediate events) are independent. Finally, the third layer allows us to return the probability value for a given formula, possibly mapping atoms to an arbitrary probability value $p$. Note that, for all three layers, we usually assign values to $e \in \mathtt{BE}$. We can however assign values to IEs if 1. $e$ is a module [20], i.e., all paths between descendants of $e$ and the rest of the FT pass through $e$ 2. and none of the descendants of $e$ are present in the formula. If so, we prune that (sub)FT and treat occurring IEs as BEs.

$$\phi ::= e \mid \neg\phi \mid \phi \wedge \phi \mid \phi[e \mapsto 0] \mid \phi[e \mapsto 1] \mid \mathrm{MCS}(\phi) \mid \mathrm{MPS}(\phi)$$

$$\psi ::= \neg\psi \mid \psi \wedge \psi \mid \Pr_{\bowtie p}(\phi \mid \phi) \mid \psi[e \mapsto q] \mid \mathrm{IDP}(\phi, \phi)$$

$$\xi ::= \Pr(\phi \mid \phi) \mid \xi[e \mapsto q]$$

where $\bowtie \in \{<, \leq, =, \geq, >\}$.

**Syntactic sugar.** We define the following derived operators where formulae $\theta$ are in the set of layer one or layer two formulae, i.e., such that $\theta \in X_1 \cup X_2$:

$$\theta_1 \vee \theta_2 ::= \neg(\neg\theta_1 \wedge \neg\theta_2) \qquad \theta_1 \not\Leftrightarrow \theta_2 ::= \neg(\theta_1 \Leftrightarrow \theta_2)$$

$$\theta_1 \Rightarrow \theta_2 ::= \neg(\theta_1 \wedge \neg\theta_2) \qquad \mathrm{SUP}(e) ::= \mathrm{IDP}(e, e_{top})$$

$$\theta_1 \Leftrightarrow \theta_2 ::= (\theta_1 \Rightarrow \theta_2) \wedge (\theta_2 \Rightarrow \theta_1)$$

$$\underset{\bowtie k}{\mathrm{Vot}}(\phi_1, \dots, \phi_N) ::= \bigvee_{\substack{U \subseteq \{1,\dots,N\} \\ |U| \bowtie k}} \left( \bigwedge_{u \in U} \phi_u \right) \wedge \left( \bigwedge_{u \in \{1,\dots,N\} \setminus U} \neg\phi_u \right)$$

with $k \leq N$.

### 3.2 Semantics

The semantics for our logic is twofold. For the first layer of PFL, formulae are evaluated on a Boolean status vector $\overline{b}$ and on a tree $T$. Atomic formulae $e$ are satisfied by $\overline{b}$ and $T$ if the structure function in Def. 2 returns 1 with these $\overline{b}$ and $e$ as input. Formally:

$$\begin{array}{ll}
\overline{b}, T \models e & \text{iff } \Phi_T(\overline{b}, e) = 1 \\
\overline{b}, T \models \neg\phi & \text{iff } \overline{b}, T \not\models \phi \\
\overline{b}, T \models \phi \wedge \phi' & \text{iff } \overline{b}, T \models \phi \text{ and } \overline{b}, T \models \phi' \\
\overline{b}, T \models \phi[e_i \mapsto 0] & \text{iff } \overline{b'}, T \models \phi \text{ with } \overline{b'} = (b'_1, \ldots, b'_n) \text{ where} \\
& b'_i = 0 \text{ and for } j \neq i \text{ we have } b'_j = b_j \\
\overline{b}, T \models \phi[e_i \mapsto 1] & \text{iff } \overline{b'}, T \models \phi \text{ with } \overline{b'} = (b'_1, \ldots, b'_n) \text{ where} \\
& b'_i = 1 \text{ and } b'_j = b_j \text{ for } j \neq i \\
\overline{b}, T \models \mathrm{MCS}(\phi) & \text{iff } \overline{b}, T \models \phi \wedge (\neg\exists \overline{b'}.\ \overline{b'} \subset \overline{b} \wedge \overline{b'}, T \models \phi) \\
\overline{b}, T \models \mathrm{MPS}(\phi) & \text{iff } \overline{b}, T \models \neg\phi \wedge (\neg\exists \overline{b'}.\ \overline{b'} \supset \overline{b} \wedge \overline{b'}, T \models \neg\phi)
\end{array}$$

With $[\![\phi]\!]_T$ we denote the *satisfaction set* of vectors for $\phi$, i.e., the set of all $\overline{b}$ that satisfy $\phi$ given $T$. Semantics for the second and third layer require the introduction of probabilities. If we consider the function $\Phi_T \colon \mathbb{B}^n \times \mathtt{E} \to \mathbb{B}$, we can devise an extension such that $\Phi_T \colon \mathbb{B}^n \times X_1 \to \mathbb{B}$, where $X_1$ is the set of layer one formulae (note that $\mathtt{E} \subseteq X_1$). With a slight abuse of notation, $\Phi_T$ will now return 1 whenever the input Boolean vector satisfies the input layer one formula. With $\phi \in X_1$, we lift the structure function to $\Phi^*_T \colon \mathit{Dist}(\mathbb{B}^n) \times X_1 \to [0,1]$, where *Dist* expresses a set of probability distributions, in a standard fashion, i.e.,

$$\Phi^*_T(\mu, \phi) = \sum \{\mu(b) \mid b \in \mathbb{B}^n \text{ for which } \Phi_T(b, \phi) = 1\}$$

We further convert probabilistic status vectors $\overline{\rho} \in [0,1]^n$ to a distribution $\mu_{\overline{\rho}} \in \mathit{Dist}(\mathbb{B}^n)$:

$$\mu_{\overline{\rho}}(b_1, \ldots, b_k) = \prod_{i=1}^{k} (b_i \times \rho_i + (1 - b_i) \times (1 - \rho_i))$$

We can then define semantics for the second syntactic layer as follows:

$$\begin{array}{ll}
\overline{\rho}, T \models \neg\psi & \text{iff } \overline{\rho}, T \not\models \psi \\
\overline{\rho}, T \models \psi \wedge \psi' & \text{iff } \overline{\rho}, T \models \psi \text{ and } \overline{\rho}, T \models \psi' \\
\overline{\rho}, T \models \Pr_{\bowtie p}(\phi \mid \phi') & \text{iff } \Phi^*_T(\mu_{\overline{\rho}}, \phi \wedge \phi') / \Phi^*_T(\mu_{\overline{\rho}}, \phi') \bowtie p \\
\overline{\rho}, T \models \psi[e_i \mapsto q] & \text{iff } \overline{\rho}[\rho_i \mapsto q], T \models \psi \\
\overline{\rho}, T \models \mathrm{IDP}(\phi, \phi') & \text{iff } \Phi^*_T(\mu_{\overline{\rho}}, \phi \wedge \phi') = \Phi^*_T(\mu_{\overline{\rho}}, \phi) \cdot \Phi^*_T(\mu_{\overline{\rho}}, \phi')
\end{array}$$

Finally, to define semantics for the third layer we let $\mathsf{Val}_{\overline{\rho}, T} \colon X_3 \to [0,1]$ define an evaluation function of layer three formulae in $X_3$:

$$\begin{array}{l}
\mathsf{Val}_{\overline{\rho}, T}(\Pr(\phi \mid \phi')) = \Phi^*_T(\mu_{\overline{\rho}}, \phi \wedge \phi') / \Phi^*_T(\mu_{\overline{\rho}}, \phi') \\
\mathsf{Val}_{\overline{\rho}, T}(\xi[e_i \mapsto q]) \ = \mathsf{Val}_{\overline{\rho}[\rho_i \mapsto q], T}(\xi)
\end{array}$$

Furthermore we write $T \models \theta$, meaning $\forall \overline{\rho}\ .\ \overline{\rho}, T \models \theta$.

# 4 Case Study: Examples

We showcase the potential of our logic by presenting two case studies: a COVID-19 related FT [4, 32] and the FT for an oil/gas pipeline [45].

## 4.1 COVID-19 FT

The TLE represents a COVID-19 infected worker on site, abbreviated *IWoS*. As shown in Fig. 2, the FT considers events in several categories: COVID-19 pathogens and reservoirs (i.e., germs and objects carrying the virus); their mode of transmissions; the presence of susceptible hosts, infected objects and workers; physical contacts as well as human errors. Note that Fig. 2 contains several repeated basic events (marked with a dashed border): *IT*, *PP*, *H1* and *IW*. This TLE *IWoS* is refined via an AND-gate with three children. Thus, for the TLE to occur the following must happen: COVID pathogens/COVID infected objects must exist, there has to be a susceptible host and COVID pathogens must be transmitted in some way to this host. These events are captured by corresponding subtrees: the **purple** OR-gate $CP/R$ refines the existence of COVID pathogens/-COVID infected objects, the OR-gate *MoT* in **teal** refines modes of transmission and the AND-gate *SH* in **orange** details the presence of a susceptible host.

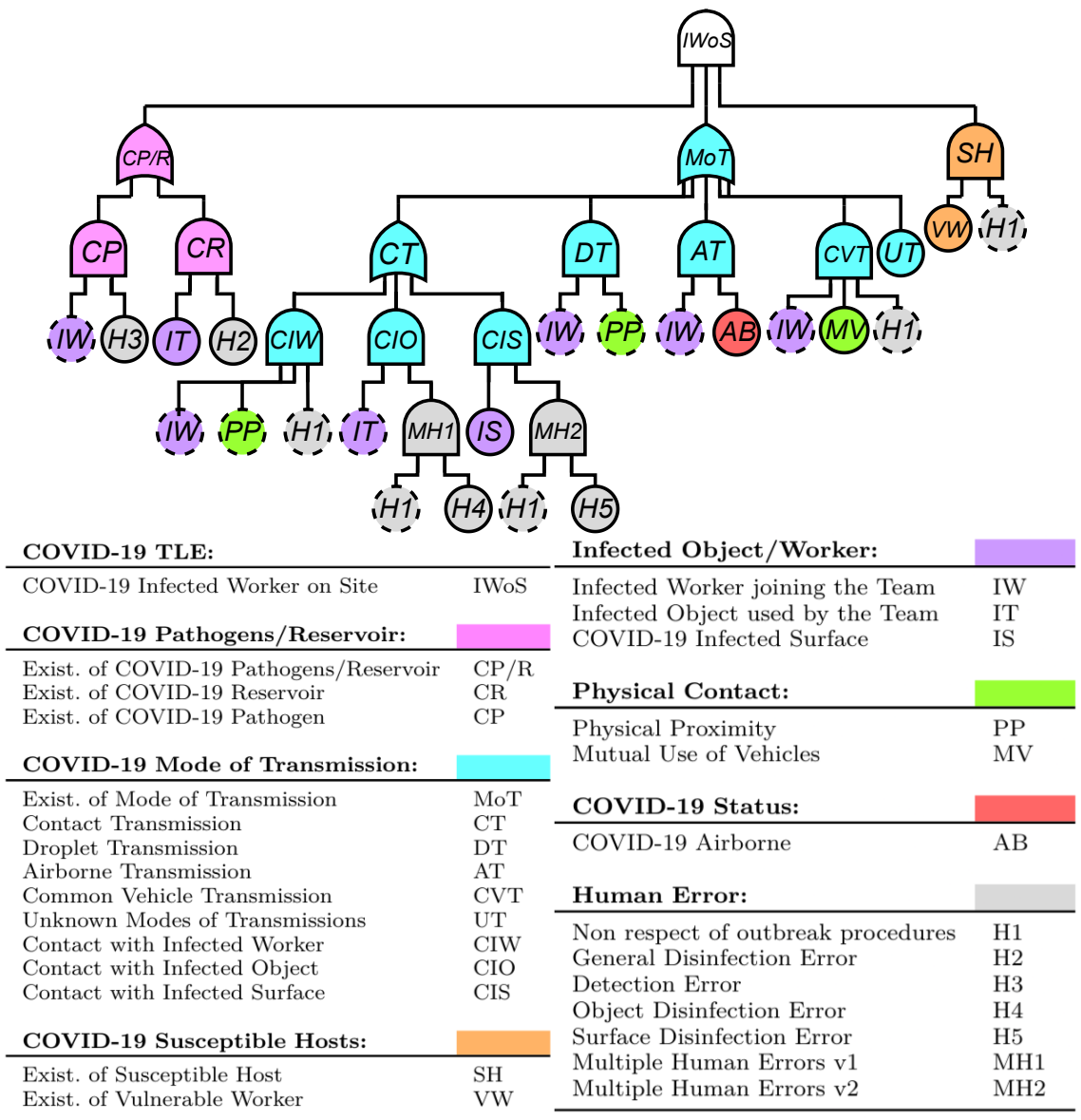

Figure 2: COVID-19 FT.

**Properties.** Following, we specify some properties using natural language and present the corresponding PFL formulae: 1) What are all the MCSs for the modes of transmission that include errors in objects and surfaces disinfection? $[\![\text{MCS}(MoT) \wedge H4 \wedge H5]\!]_T$; 2) Is the probability of TLE smaller than 0.03, if physical proximity occurred? $\text{Pr}_{\leq 0.03}(IWoS)[PP \mapsto 1]$; 3) Assume that the probability of an infected worker on the team equals 0.25. How does that affect the probability of TLE? $\text{Pr}(IWoS)[IW \mapsto 0.25]$; 4) Assume that both COVID-19 pathogens and a vulnerable worker exist. Does this imply that $P(IWoS) \geq 0.15$? $\text{Pr}_{=1}(CP) \wedge \text{Pr}_{=1}(VW) \Rightarrow \text{Pr}_{\geq 0.15}(IWoS)$.

## 4.2 Oil/gas pipeline FT

The TLE represents the failure of an oil/gas pipeline, abbreviated $O/GPF$. As shown in Fig. 3, the FT considers events in several categories: failures like ruptures and punctures; third party interference; different kinds of corrosion; incorrect performance of some operations (e.g., maintenance); unreasonable design choices; as well as defects on pipes. Fig. 3 contains several repeated basic events (again, marked with a dashed border): $WW$, $H_2S$, $O_2$, $CO_2$ and $IAC$. Furthermore, multiple sub-trees are referenced/repeated in different places: those are marked using labelled triangles. The TLE $O/GPF$ is refined via an OR-gate with two children, in blue. Thus, for the TLE to occur either a rupture or a puncture must happen. These two events are captured by corresponding subtrees. The rupture subtree (top-right of Fig. 3) is refined by an OR-gate with six children: the green OR-gate $TPI$ refines possible interference by third parties; the violet OR-gate $Cor$ refines modes of pipes corrosion; the yellow subtree $B$ refines modes in which pipes could be defective; the dove gray OR-gate $IO$ details possible incorrect operations; the lime green OR-gate $UD$ details unreasonable design choices; and the pink OR-gate $GH$ refines possible geological hazards. Similarly, the puncture subtree (bottom of Fig. 3) is refined by an OR-gate with two children: the orange OR-gate $CoT$ refines modes in which corrosion can make pipes thinner — with a detailed subtree in light blue refining medium corrosion; and the OR-gate $DoP$ in yellow that refines modes in which pipes could be defective.

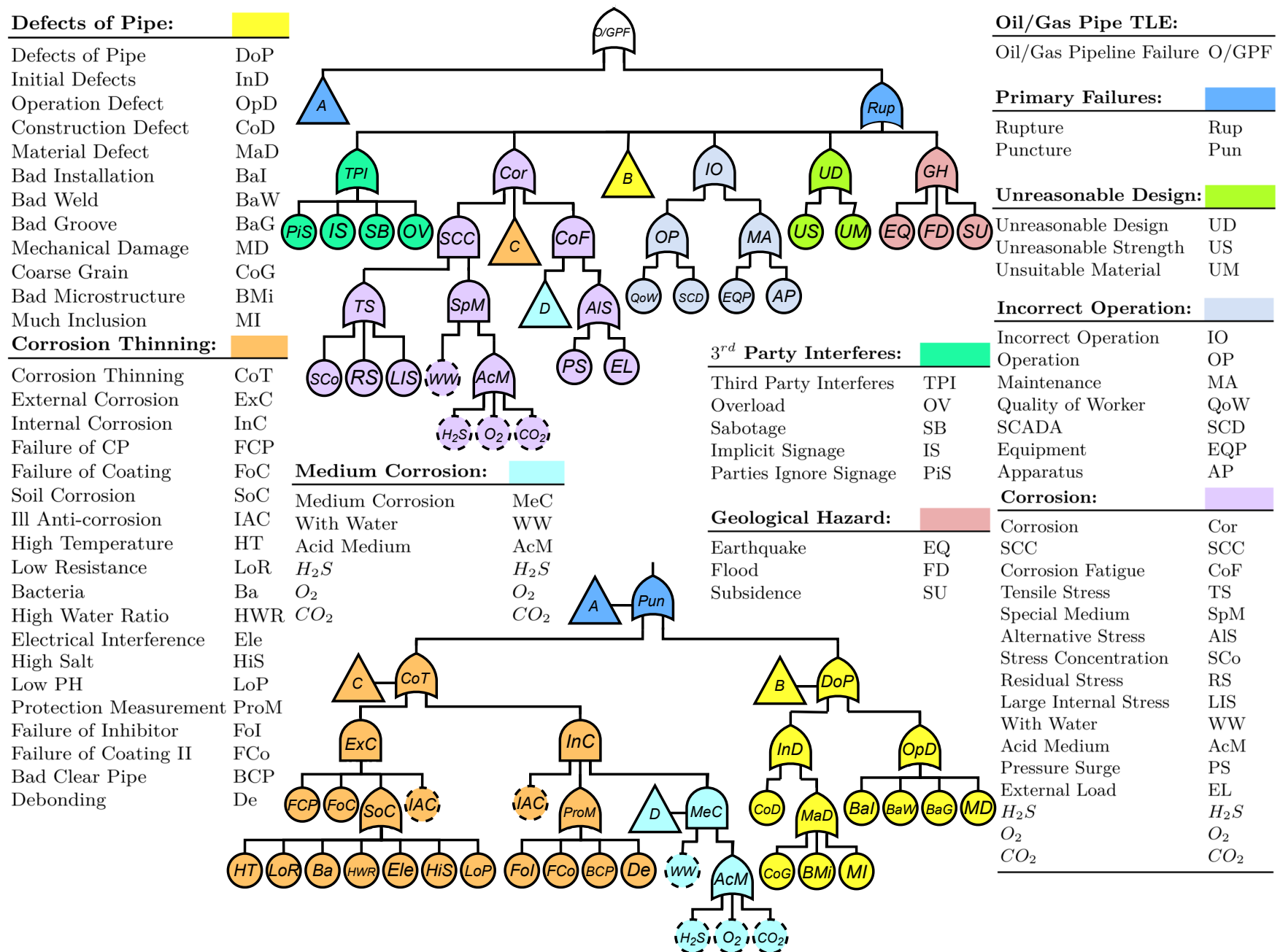

Figure 3: Oil/Gas Pipeline FT.

**Properties.** We specify some properties using natural language and present the corresponding PFL formulae: 1) What are all the MPSs for pipes rupture that include the absence of water as a corrosive medium, $H_2S$, $O_2$ and $CO_2$? $[\![\mathrm{MPS}(Rup) \land \neg WW \land \neg H_2S \land \neg O_2 \land \neg CO_2]\!]_T$; 2) Assume that $H_2S$ shows up in the pipes with 0.25% probability. What is the probability of pipes corrosion, given that corrosion happens with water with 2% probability and that pressure surges with 1% probability? $\Pr(\mathit{Cor})[\mathit{H_2S} \mapsto 0.0025, \mathit{WW} \mapsto 0.02, \mathit{PS} \mapsto 0.01]$; 3) Assume that the probability of pipes corrosion with acid is equal to 0.005. Assume also that pipes present defects in their construction material with 0.2% probability. Is the probability of TLE happening lower than 1.2%? $\Pr_{\leq 0.012}(\mathit{O/GPF})[\mathit{AcM} \mapsto 0.005, \mathit{MaD} \mapsto 0.02]$.

## 5 LangPFL: A Domain Specific Language for PFL

**Design of LangPFL.** To ease usability of PFL, we present LangPFL, a Domain Specific Language (DSL) to specify properties in PFL. The strive for a simple way to specify properties involving probability on FT was a specific need we uncovered while conducting interviews with a domain expert from industry [3]. Defining languages and tools for properties and requirements specification is common practice. In [17] the authors capture high-level requirements in a human readable form by presenting SADL, a controlled English requirements capture language, alongside its tool suite ASSERT. Other controlled natural languages for knowledge representation include Processable English (PENG) [44], Controlled English to Logic Translation (CELT) [34] and Computer Processable Language (CPL) [13]. LangPFL is inspired by these languages for their ease of use and close proximity to natural language. Finally, another notable example is FRETish [15], a structured natural language used to specify requirements and to translate them into Linear Temporal Logic (LTL), developed at NASA and supported by the FRET tool [23]. Other than for its usability, FRETish inspired us with the clear way in which the scope, conditions and component of specified properties are clearly separated from desired behaviours on timing and responses. LangPFL expresses only a fragment of PFL: most notably, nesting of formulae is disallowed. By doing so, we retain most of the expressiveness of PFL while making property specification easier. FT elements are referred to with their label and each operator in PFL has a counterpart in the DSL: Boolean operators, not, and, or, impl ...; setting the value of FT elements to Boolean or probability values, set, setp; MCSs and MPSs, MCS[...], MPS[...]; operators to check probability thresholds/compute probability values, P[...] $\bowtie$ ..., P[...]; and to check for independence between FT elements IDP[..., ...].

| Natural Language | PFL | LangPFL |
|---|---|---|
| What are all the MCSs for the modes of transmission that include errors in objects and surfaces disinfection? | $[\![\mathrm{MCS}(\mathit{MoT}) \land \mathit{H4} \land \mathit{H5}]\!]_T$ | **assume:**<br>**computeall:**<br>MCS[MoT] and<br>H4 and H5 |
| Is the probability of TLE smaller than 0.03, if physical proximity occurred? | $\Pr_{\leq 0.03}(\mathit{IWoS})[\mathit{PP} \mapsto 1]$ | **assume:**<br>setp PP = 1<br>**check:**<br>P[IWoS] $\leq$ 0.03 |
| Assume that the probability of an infected worker on the team equals 0.25. How does that affect the probability of TLE? | $\Pr(\mathit{IWoS})[\mathit{IW} \mapsto 0.25]$ | **assume:**<br>setp IW = 0.25<br>**compute:**<br>P[IWoS] |
| Assume that both COVID-19 pathogens and a vulnerable worker exist. Does this imply that $P(\mathit{IWoS}) \geq 0.15$? | $\Pr_{=1}(\mathit{CP}) \land \Pr_{=1}(\mathit{VW}) \Rightarrow \Pr_{\geq 0.15}(\mathit{IWoS})$ | **assume:**<br>setp CP = 1<br>setp VW = 1<br>**check:**<br>P[IWoS] $\geq$ 0.15 |
| What are all the MPSs for pipes rupture that include the absence of water as a corrosive medium, $\mathit{H_2S}$, $\mathit{O_2}$ and $\mathit{CO_2}$? | $[\![\mathrm{MPS}(\mathit{Rup}) \land \neg \mathit{WW} \land \neg \mathit{H_2S} \land \neg \mathit{O_2} \land \neg \mathit{CO_2}]\!]_T$ | **assume:**<br>**computeall:**<br>MPS [Rup] and<br>not WW and<br>not $\mathit{H_2S}$<br>and not $\mathit{O_2}$<br>and not $\mathit{CO_2}$ |
| Assume that $\mathit{H_2S}$ shows up in the pipes with 0.25% probability. What is the probability of pipes corrosion, given that corrosion happens with water with 2% probability and that pressure surges with 1% probability? | $\Pr(\mathit{Cor})[\mathit{H_2S} \mapsto 0.0025, \mathit{WW} \mapsto 0.02, \mathit{PS} \mapsto 0.01]$ | **assume:**<br>setp $\mathit{H_2S}$ = 0.0025<br>setp WW = 0.02<br>setp PS = 0.01<br>**compute:**<br>P[Cor] |
| Assume that the probability of pipes corrosion with acid is equal to 0.005. Assume also that pipes present defects in their construction material with 0.2% probability. Is the probability of TLE lower than 1.2%? | $\Pr_{\leq 0.012}(\mathit{O/GPF})[\mathit{AcM} \mapsto 0.005, \mathit{MaD} \mapsto 0.02]$ | **assume:**<br>setp AcM = 0.005<br>setp MaD = 0.02<br>**check:**<br>P[O/GPF]$\leq$0.012 |

Table 1: Properties: natural language, PFL, LangPFL.

LangPFL **Templates.** Properties can be specified in LangPFL by utilizing operators inside structured templates. Assumptions on the status of FT elements can be specified under the **assume** keyword. These assumptions will be automatically integrated in the translated formula accordingly, e.g., set or setp will be translated with the according operators to set evidence, while other assumptions will be the antecedent of an implication. A second keyword separates specified formulae from the assumptions and dictates the desired result: **compute** and **computeall** compute and return desired values, i.e., probability values and lists of MCSs/MPSs respectively, while **check** establishes if a specified property holds.

**Case studies.** In Table 1 we showcase the properties specified in Sec. 4 and their respective translation in LangPFL.

## 6 Model Checking Algorithms

**Overview.** With PFL extending previous work [32], algorithms to compute satisfiability of layer one formulae remain unchanged. In particular, it is possible to model check PFL over a FT and a Boolean vector $\overline{b}$ when considering layer one formulae. Furthermore, we can collect all Boolean vectors $\overline{b}$ such that $\overline{b}, T \models \phi$. As noted in [32], checking if $\overline{b}, T \models \phi$ holds is trivial if $\phi$ is a layer one formula that does not contain an MCS or MPS operator. In that case, we can simply substitute the values of $\overline{b}$ in $\phi$ and see if the Boolean expression evaluates to true. This holds true also when considering a given probabilistic vector $\overline{\rho}$, a tree-shaped FT and a layer two/three formula $\theta$ that does not contain operators for MCS or MPS. In this case, values can be computed following usual probability laws. For the other cases, the computation becomes more complex, and procedures involving binary decision diagrams (BDDs) are necessary. Algorithms for the Boolean scenarios are described in Appendix A.3 and Appendix A.4 respectively. When reasoning about satisfiability of second layer formulae, algorithms present differences. As such, we present three novel algorithms for PFL: 1. Given a vector $\overline{\rho}$, a FT $T$ and a formula $\psi$, check if $\overline{\rho}, T \models \psi$ (Sec. 6.4), 2. Given $T$ and $\psi$, compute regions of the parameter space where $T \models \psi$ (Sec. 6.5), 3. Given a FT $T$ and a formula $\psi$, check whether $T \models \psi$ for all $\overline{\rho}$ (Sec. 6.6). In continuity with previous work, all three algorithms are based on construction and manipulation of BDDs: first, FT elements that appear in a given layer one formula are identified. Then, BDDs for these elements are selectively constructed (see

Algo. 5) and stored to reduce computation time. Finally, these BDDs are manipulated and equipped with probabilities (see Algo. 1) to reflect semantics of the operators of PFL. This translation to BDDs constitutes a formal ground that permits to address these procedures in a uniform way, while integrating novel work presented in this paper with previous algorithms. A brief overview of each algorithm is given in Sec. 6.4, Sec. 6.5 and Sec. 6.6 respectively.

### 6.1 (Reduced Ordered) Binary Decision Diagrams

BDDs are directed acyclic graphs (DAGs) that compactly represent Boolean functions [2] by reducing redundancy. Depending on variable's ordering, BDD's size can grow linearly in the number of variables and at worst exponentially. In practice, BDDs are heavily used, including in FT analysis and in their security-related counterpart, attack trees (ATs) [37, 12]. Formally, a BDD is a rooted DAG $\mathbf{B}_f$ that represents a Boolean function $f\colon \mathbb{B}^n \to \mathbb{B}$ over variables $\mathit{Vars} = \{x_i\}_{i=1}^n$. Each nonleaf $w$ has two outgoing arrows, labeled $\mathtt{0}$ and $\mathtt{1}$, and a label $\mathit{Lab}(w) \in \mathit{Vars}$; furthermore, each leaf has a label $\mathtt{0}$ or $\mathtt{1}$. Given a $b$ in $\mathbb{B}^n$, the BDD is used to compute $f(b)$ as follows: starting from the top, upon arriving at a node w with $\mathit{Lab}(w) = x_i$, one takes the $\mathtt{0}$-edge if $b_i = 0$ and the $\mathtt{1}$-edge if $b_i = 1$. The label of the leaf one ends up in, is then equal to $f(b)$. A function $f$ can be represented by multiple BDDs, but has a unique *reduced ordered* representative, or ROBDD [5, 10], where the $x_i$ occur in ascending order, and the BDD is reduced as much as possible by removing irrelevant nodes and merging duplicates. This is formally defined below; we let $\mathit{Low}(w)$ (resp. $\mathit{High}(w)$) be the endpoint of $w$'s $\mathtt{0}$-edge (resp. $\mathtt{1}$-edge) and let $R_{\mathbf{B}}$ be the BDD root.

**Definition 5** (***Reduced Ordered Binary Decision Diagram ((RO)BDD)***). *Let Vars be a set. A* BDD *over Vars is a tuple* $\mathbf{B} = (W, A, \mathit{Lab}, u)$ *where* $(W, A)$ *is a rooted directed acyclic graph, and* $\mathit{Lab}\colon W \to \mathit{Vars} \sqcup \{\mathtt{0}, \mathtt{1}\}, u\colon A \to \{\mathtt{0}, \mathtt{1}\}$ *are maps such that: 1. Every nonleaf* $w$ *has exactly two outgoing edges* $a, a'$ *with* $u(a) \neq u(a')$*, and* $\mathit{Lab}(w) \in \mathit{Vars}$*; 2. Every leaf* $w$ *has* $\mathit{Lab}(w) \in \{\mathtt{0}, \mathtt{1}\}$*. 3. Vars are equipped with a total order,* $\mathbf{B}_f$ *is thus defined over a pair* $\langle \mathit{Vars}, < \rangle$*; 4. the variable of a node is of lower order than its children, that is:* $\forall\, w \in W_n \,.\, \mathit{Lab}(w) < \mathit{Lab}(\mathit{Low}(w)), \mathit{Lab}(\mathit{High}(w))$*; 5. the children of non terminal nodes are distinct nodes; 6. nodes are uniquely determined by their label, low child and high child.*

### 6.2 Translating FTs/formulae to BDDs

**Translations.** We shortly sketch the idea of translating a layer one formula and a (sub)tree to BDDs. As mentioned, to translate formulae to BDDs, FT elements that appear in a given formula are identified. Then, BDDs representing these elements are selectively constructed and stored to reduce computation time. Finally, operations on these BDDs are performed to reflect semantics of the operators in PFL.

**Translating FTs to BDDs.** As a first step, a translation from FTs to BDDs is needed [32]. These BDDs represent exactly the structure function of (sub)trees. In the following paragraphs we assume $\mathit{Vars} = \mathtt{V} \dot\cup \mathtt{V}'$, where the set of variables $\mathtt{V} = \mathtt{BE}$ and the set of primed variables $\mathtt{V}' = \{e' | e \in \mathtt{BE}\}$ (used for the BDD translation of the MCS operator, see Appendix A.2). Furthermore, we keep $\mathit{Var}_{\mathbf{B}}\colon \mathtt{BDD} \to \mathit{Vars}$ to be a function that returns variables occurring in a BDD [32]. Then, our translation function $\Psi_{FT}\colon \mathtt{E} \to \mathtt{BDD}$ takes elements of a FT as input and maps them to BDDs. For an exact definition of $\Psi_{FT}$ see Appendix A.1.

**Translating formulae.** With BDDs for FTs, the next step consists in manipulating them to mirror PFL operators in layer one. I.e., given $\Psi_{FT}$ and a FT $T$, for every PFL formula $\phi$ in the set of PFL layer one formulae $X_1$ there exists a translation to BDDs $\underline{\mathbf{B}}_T\colon X_1 \to \mathtt{BDD}$ in Algo. 5 (see Appendix A.2). The implementation of this procedure abides the dynamic programming standards: by caching, we would reuse the translation of (sub)trees and (sub)formulae between different analyses without recomputing them each time anew.

### 6.3 Equipping BDDs with probabilities

Once we obtain BDDs for FTs/$\phi$-formulae, we can construct a function $\Phi^*_T(\mu_{\overline{x}}, \phi)$ from $[0,1]^n$ to $[0,1]$ that computes the probability value of $\phi$ given probability values in $\overline{x}$, where $\overline{x}$ can be substituted with any $\overline{\rho}$. Algo. 1 shows this procedure: first, we compute $\underline{\mathbf{B}}_T(\phi)$ via Algo. 5, we then obtain a polynomial $\mathit{poly}(\underline{\mathbf{B}}_T(\phi))$ representing $\underline{\mathbf{B}}_T(\phi)$ via $\mathsf{value}(R_{\underline{\mathbf{B}}_T(\phi)})$, where $\mathsf{value}(w_i \not\in W_t) = (1 - x_i) \cdot \mathsf{value}(\mathit{Low}(w_i)) + x_i \cdot \mathsf{value}(\mathit{High}(w_i))$, $\mathsf{value}(\top) = 1$ and $\mathsf{value}(\bot) = 0$. $x_1, \ldots, x_n \in \overline{x}$ are parameters of the constructed function and can be substituted in $\mathit{poly}(\underline{\mathbf{B}}_T(\phi))$ with values from an arbitrary $\overline{\rho}$ to compute the overall probability value of the BDD for FT/$\phi$-formula.

**Algorithm 1** Obtain $\Phi^*_T(\mu_{\overline{x}}, \phi)$ for $\underline{\mathbf{B}}_T(\phi)$.

**Input**: FT $T$, formula $\phi$
**Output:** function $\Phi^*_T(\mu_{\overline{x}}, \phi)\colon [0,1]^n \to [0,1]$
where $x_1, \ldots, x_n \in \overline{x}$ are function parameters
**Method:**
$\underline{\mathbf{B}}_T(\phi) \leftarrow$ Algo. 5$(T, \phi)$
$\mathit{poly}(\underline{\mathbf{B}}_T(\phi)) \leftarrow \mathsf{value}(R_{\underline{\mathbf{B}}_T(\phi)})$, where:
- $\mathsf{value}(w_i \not\in W_t) = (1 - x_i) \cdot \mathsf{value}(\mathit{Low}(w_i))$ $+ x_i \cdot \mathsf{value}(\mathit{High}(w_i))$
- $\mathsf{value}(\top) = 1$ and $\mathsf{value}(\bot) = 0$

**return** $\mathit{poly}(\underline{\mathbf{B}}_T(\phi))$

### 6.4 Algorithm 2: Model checking PFL over a FT and a $\overline{\rho}$

**Overview.** Given a specific vector $\overline{\rho}$, a FT $T$ and a PFL layer two formula $\psi$, we want to check if $\overline{\rho}, T \models \psi$. To do so, if we come across a layer one formula $\phi$ we translate it to a BDD, we equip the resulting BDD with probabilities obtained from $\overline{\rho}$ and we compute whether the resulting value respects the threshold set in the given layer two formula $\psi$. Boolean connectives are resolved as usual and independence is checked according to probability laws once the value for the respective BDD is computed. For the corresponding layer three formulae $\xi$, we would simply return the value computed from the BDD instead of comparing it to the given layer two threshold.

**Algo. 2.** This algorithm shows a procedure to check if $\overline{\rho}, T \models \psi$, given $\overline{\rho}$, $T$ and $\psi$. Boolean connectives are handled as usual via case distinction. In the same way, probability values in $\overline{\rho}$ are replaced by mappings in $\psi$, if any. For $\mathrm{Pr}_{\bowtie p}(\phi \mid \phi')$, we compute the BDD $\underline{\mathbf{B}}_T(\phi_n)$ for each $\phi_n$ of the respective layer one formulae via Algo. 1. Finally, we compute the conditional probability $P(\phi \mid \phi')$. If the returned value respects the threshold set in $\psi$ we return *True*, *False* otherwise. For IDP we follow an analogous procedure: we compute probability values of needed layer one inner formulae and we return *True* if they are stochastically independent. An algorithm for layer three formulae $\xi$ would simply return the conditional probability value for $\Pr(\phi \mid \phi')$, after potentially modifying $\overline{\rho}$ and computing $P(\underline{\mathbf{B}}_T(\phi_n))$.

**Example.** Let us consider the subtree in Fig. 1 and a vector with probability values for $WW$, $H_2S$, $O_2$ and $CO_2$ respectively: $\overline{\rho} = (0.002, 0.001, 0.0015, 0.002)$. Suppose we want to know if $P(MeC)$ is lower or equal to 0.0001, assuming the scenario where $P(H_2S) = 0.0023$ and $P(WW) = 0.015$, i.e., formally with $\psi = \Pr_{\leq 0.0001}(MeC)[H_2S \mapsto 0.0023, WW \mapsto 0.015]$. First, $\overline{\rho}$ would be modified as per the new assignments in $\psi$: $\overline{\rho} = (0.015, 0.0023, 0.0015, 0.002)$. Then, Algo. 2 is called again with the modified $\overline{\rho}$ and the BDD $\underline{\mathbf{B}}_T(MeC)$ for $MeC$ is constructed (see Fig. 4). The value for the BDD is computed via Algo. 1. The result (0.000087) is lower than the threshold in $\psi$, the formula is satisfied and the algorithm returns *True*.

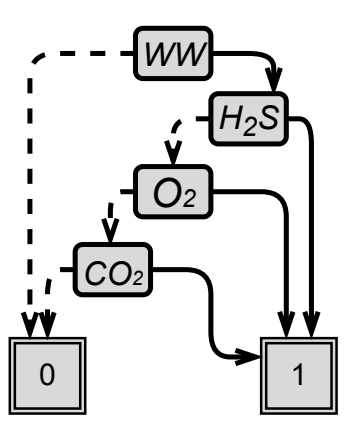

Figure 4: BDD for Fig. 1.

---

**Algorithm 2** Check if $\overline{\rho}, T \models \psi$, given $\overline{\rho}$, $T$ and $\psi$.

---

**Input**: prob. vector $\overline{\rho}$, FT $T$, formula $\psi$
**Output:** *True* iff $\overline{\rho}, T \models \psi$, *False* otherwise.
**Method:**
**if** $\psi = \neg\psi'$ **then return not**(Algo. 2$(\overline{\rho}, T, \psi')$)
**else if** $\psi = \psi' \wedge \psi''$ **then return** Algo. 2$(\overline{\rho}, T, \psi')$ **and** Algo. 2$(\overline{\rho}, T, \psi'')$
**else if** $\psi = \Pr_{\bowtie p}(\phi \mid \phi')$ **then**
  $P(\underline{\mathbf{B}}_T(\phi \wedge \phi')), P(\underline{\mathbf{B}}_T(\phi')) \leftarrow$Algo. 1$(T, \phi \wedge \phi')(\overline{\rho})$, Algo. 1$(T, \phi')(\overline{\rho})$
  $P(\phi \mid \phi') = \frac{P(\underline{\mathbf{B}}_T(\phi \wedge \phi'))}{P(\underline{\mathbf{B}}_T(\phi'))}$
  **return** $P(\phi \mid \phi') \bowtie p$
**else if** $\psi = \psi'[e_i \mapsto q]$ **then return** Algo. 2$(\overline{\rho}[\rho_i \mapsto q], T, \psi')$
**else if** $\psi = \mathrm{IDP}(\phi, \phi')$ **then**
  $P(\underline{\mathbf{B}}_T(\phi)), P(\underline{\mathbf{B}}_T(\phi')), P(\underline{\mathbf{B}}_T(\phi \wedge \phi')) \leftarrow$Algo. 1$(T, \phi)(\overline{\rho})$, Algo. 1$(T, \phi')(\overline{\rho})$, Algo. 1$(T, \phi \wedge \phi')(\overline{\rho})$
  **return** $P(\underline{\mathbf{B}}_T(\phi)) \cdot P(\underline{\mathbf{B}}_T(\phi')) = P(\underline{\mathbf{B}}_T(\phi \wedge \phi'))$
**end if**

---

### 6.5 Algorithm 3: Computing regions where $\psi$-formulae are satisfied

**Overview.** Given a FT $T$ and a layer two formula $\psi$, we want to find the region $S_{\text{yes}}$ in $[0, 1]^n$ of all $\overline{\rho}$ that satisfy $\psi$. Typically, such a region is defined by large polynomials, and therefore difficult to describe analytically. Instead, we provide an algorithm that approximates this region up to a given level of precision. Such an approximation is given in the definition below: it consists of a region $S_{\text{yes}}$ where $\psi$ is known to hold, a region $S_{\text{no}}$ where $\psi$ does not hold, and the remainder $S_{\text{maybe}}$ is of limited volume.

**Definition 6.** *Let $T$ be a FT, let $\varepsilon \in (0,1]$, and let $\psi$ be a layer two formula. A* $\varepsilon$-partition for $\psi$ *is a partition $(S_{yes}, S_{no}, S_{maybe})$ of $[0,1]^n$ such that: 1. $\overline{\rho}, T \models \psi$ for all $\overline{\rho} \in S_{yes}$; 2. $\overline{\rho}, T \not\models \psi$ for all $\overline{\rho} \in S_{no}$; 3.* $\mathrm{Vol}(S_{maybe}) \leq \varepsilon$*, where* Vol *denotes $n$-dimensional volume.*

---

**Algorithm 3** Given $T$, find $\varepsilon$-partition for $\Pr_{\geq p}(\phi|\phi')$.

---

**Input**: FT $T$, formulae $\phi, \phi'$, reals $p, \varepsilon \in (0,1]$.
**Output:** $\varepsilon$-partition $(S_{\text{yes}}, S_{\text{no}}, S_{\text{maybe}})$ for $\Pr_{\geq p}(\phi|\phi')$.
**Method:**
$\mathscr{B}_{\text{maybe}} \leftarrow \{[0,1]^n\}$; $V_{\text{maybe}} \leftarrow 1$; $S_{\text{yes}}, S_{\text{no}} \leftarrow \varnothing$
**while** $V_{\text{maybe}} > \varepsilon$ **do**
  Pick $B = \prod_{i=1}^n [l_i, u_i]$ from $\mathscr{B}_{\text{maybe}}$ with maximal volume
  $\mathscr{B}_{\text{maybe}} \leftarrow \mathscr{B}_{\text{maybe}} \setminus \{B\}$
  $V_{\text{maybe}} \leftarrow V_{\text{maybe}} - \mathrm{Vol}(B)$
  $\mathscr{B}_{\text{test}} \leftarrow \left\{ \prod_{i=1}^n I_i \;\middle|\; \forall i. I_i \in \{[l_i, \frac{l_i+u_i}{2}], [\frac{l_i+u_i}{2}, u_i]\} \right\}$
  **for each** $B' = \prod_{i=1}^n [l'_i, u'_i] \in \mathscr{B}_{\text{test}}$ **do**
    $A \leftarrow \{\overline{\rho} \in [0,1]^n \mid \forall i. \rho_i \in \{l'_i, u'_i\}\}$
    $p_{\min} \leftarrow \min_{\overline{\rho} \in A} \frac{\text{Algo. 1}(T, \phi \wedge \phi')(\overline{\rho})}{\text{Algo. 1}(T, \phi')(\overline{\rho})}$
    $p_{\max} \leftarrow \max_{\overline{\rho} \in A} \frac{\text{Algo. 1}(T, \phi \wedge \phi')(\overline{\rho})}{\text{Algo. 1}(T, \phi')(\overline{\rho})}$
    **if** $p \leq p_{\min}$ **then** $S_{\text{yes}} \leftarrow S_{\text{yes}} \cup B'$
    **else if** $p > p_{\max}$ **then** $S_{\text{no}} \leftarrow S_{\text{no}} \cup B'$
    **else** $\mathscr{B}_{\text{maybe}} \leftarrow \mathscr{B}_{\text{maybe}} \cup \{B'\}$; $V_{\text{maybe}} = V_{\text{maybe}} + \mathrm{Vol}(B')$
    **end if**
  **end for**
**end while**
$S_{\text{maybe}} \leftarrow \bigcup \mathscr{B}_{\text{maybe}}$
**return** $(S_{\text{yes}}, S_{\text{no}}, S_{\text{maybe}})$

---

**Algo. 3.** An algorithm finding a $\varepsilon$-partition for formulae of the form $\psi = \Pr_{\geq p}(\phi|\phi')$ is given in Algo. 3; it works as follows. We have a set $\mathscr{B}_{\text{maybe}}$ of candidate hypercubes, which starts as the singleton $\{[0,1]^n\}$. One by one, we take hypercubes $B$ from $\mathscr{B}_{\text{maybe}}$, and divide them into $2^n$ smaller hypercubes. For each of the smaller hypercubes $B'$, we check whether $\overline{\rho}, T \models \psi$ for all $\psi \in B'$; if so, we add $B'$ to $S_{\text{yes}}$. If $\overline{\rho}, T \not\models \psi$ for all $\psi \in B'$, we add $B'$ to $S_{\text{no}}$. If neither is true, then we add $B'$ to $\mathscr{B}_{\text{maybe}}$, so that later it is split up again. The algorithm stops when the joint volume of all hypercubes in $\mathscr{B}_{\text{maybe}}$ is at most $\varepsilon$. Literature in the area of parametric model checking explored this technique, also w.r.t. Markov decision processes (MDPs) [18, 22, 24, 27, 28]. However, we leverage the specific situation presented here to devise a less generic but more convenient algorithm. In fact, to check $\forall \overline{\rho} \in B'. \overline{\rho}, T \models \psi$, we use Theorem 1 (proof in Appendix B.1), which says that the minimum of $\frac{\Phi^*_T(\overline{\rho}, \phi \wedge \phi')}{\Phi^*_T(\overline{\rho}, \phi')}$ $\left(\text{computed as } \frac{\text{Algo. 1}(T, \phi \wedge \phi')(\overline{\rho})}{\text{Algo. 1}(T, \phi')(\overline{\rho})}\right)$ on $B'$ is attained at one of its vertices. This means that we only need to check whether $\overline{\rho}, T \models \psi$ for the set $A$ of vertices of $B'$. The same holds for checking $\forall \overline{\rho} \in B'. \overline{\rho}, T \not\models \psi$.

**Theorem 1.** *Let $\phi, \phi'$ be layer one formulae, and let $B \subseteq [0,1]^n$ be a hyperrectangle. Then $\frac{\Phi^*_T(\overline{\rho}, \phi \wedge \phi')}{\Phi^*_T(\overline{\rho}, \phi')}$ attains its minimum and maximum (as a function of $\overline{\rho}$) at one of the vertices of $B$.*

So far, we have assumed $\psi = \mathrm{Pr}_{\geq p}(\phi|\phi')$. Formulae of the form $\mathrm{Pr}_{=p}(\phi|\phi')$ and $\mathrm{IDP}(\phi, \phi')$ generally define hypersurfaces in $[0,1]^n$ rather than regions; these can be approximated by considering the set $S_{\mathrm{maybe}}$ of a $\varepsilon$-partition, which forms an open neighborhood of the actual hypersurface. Furthermore, one finds regions for $\neg\psi$ and $\psi \wedge \psi'$ by considering complements and intersections, respectively.

### 6.6 Algorithm 4: Checking PFL $\psi$-formulae over a FT for all $\overline{\rho}$

**Overview.** Given a FT $T$ a layer two formula $\psi$, we want to check if $T \models \psi$ for all $\overline{\rho}$. In this section we discuss two different approaches to answer this question, one derived from Algo. 3 and one employing SAT solving.

**Algo. 4.** Leveraging Algo. 3, one could check whether $T \models \psi$ for all $\overline{\rho}$ by checking the parameter space in order to show that the negated formula $\neg\psi$ is unsatisfiable. If, on the other hand, we manage to find a candidate hypercube $B'$ from $\mathscr{B}_{\mathrm{maybe}}$ such that $\forall \overline{\rho} \in B'.\overline{\rho}, T \models \neg\psi$ then we can exhibit a region that serves as a counterexample for our initial question. This procedure would be bound to approximate to a given level of precision, as previously discussed.
The second possibility is to resort to SMT solving. Again, our aim is to check if the negation of the given formula is unsatisfiable. First, we translate each of the inner $\phi_n$ layer one formulae (e.g., inside $\mathrm{Pr}_{\bowtie p}(\phi \mid \phi')$ or $\mathrm{IDP}(\phi, \phi')$ operators) to BDDs, to then obtain representations of these BDDs as polynomials (see Algo. 1). By comparing these to bounds set in $\mathrm{Pr}_{\bowtie p}(\phi \mid \phi')$ operators or to the semantics of $\mathrm{IDP}(\phi, \phi')$, one can represent the original negated formula $\neg\psi$ via (in)equalities between polynomials. We then use already available SMT solvers - such as SMT-RAT [16] - as a black box to handle such an encoding. If the input representation is satisfiable, the SMT solver returns an assignment of variables to values, i.e., a counterexample probability vector for our original question.

## 7 Conclusion and Future Work

**Conclusion.** We presented PFL, a probabilistic logic for FTs that enables the construction of complex queries that capture many relevant scenarios. Furthermore, we introduced LangPFL, a domain specific language for PFL to ease property specification. We showcased their usefulness with an application of PFL and LangPFL to a COVID19-related FT and to a FT for an oil/gas pipeline. Specified properties can then be checked via the model checking algorithms, that we presented alongside relevant theorems.

**Future work.** Our work opens several relevant perspectives for future research. First, it would be interesting to extend PFL to consider timed behaviours to further extend quantitative analysis capabilities. Secondly, it would be possible to extend PFL in order to consider dynamic gates in FTs. This further validates our first point: to handle dynamic gates in dynamic FTs it would be very natural to have a logic that can express temporal properties, moving more in the direction of LTL [35] or CTL [14] or their timed variants TLTL [36] and TCTL [1]. Moreover, it is foreseeable to extend the proposed framework to security variants of FTs, attack trees (ATs) [7, 12, 31, 39], and to their combinations, e.g., attack-fault trees (AFTs) [29]. Lastly, developing an implementation of this logic could further propel usability of PFL and LangPFL by providing hands-on feedback from domain experts acquainted with FTA.

# A Appendix: Algorithms and additional definitions for layer one formulae

Following, operations between BDDs are represented by **bold** operands e.g., $\boldsymbol{\wedge}, \boldsymbol{\vee}$. Algorithms to conduct these operations on BDDs can be found in [2, 5]. Given a set of variables $\mathtt{V} = \{v_1, \ldots, v_n\}$, existential quantification (needed to translate part of the semantics of MCS operator) can be defined as follows: $\exists v.\mathbf{B} = \textsc{Restrict}(\mathbf{B}, v, 0) \boldsymbol{\vee} \textsc{Restrict}(\mathbf{B}, v, 1)$; $\exists V.\mathbf{B} = \exists v_1.\exists v_2.\ \ldots\ \exists v_n.\mathbf{B}$.

## A.1 Translating FTs to BDDs

$\Psi_{FT}$ is defined as follows:

**Definition 7.** *The translation function of a FT T is a function* $\Psi_{FT_T}\colon \mathtt{E} \to \mathtt{BDD}$ *that takes as input an element* $e \in \mathtt{E}$. *With* $e' \in ch(e)$, *we can define* $\Psi_{FT_T}$*:*

$$\Psi_{FT_T}(e) = \begin{cases} \overline{\mathbf{B}}(e) & \text{if } e \in \mathtt{BE} \\ \bigvee \Psi_{FT_T}(e') & \text{if } e \in \mathtt{IE} \text{ and } t(e) = \mathtt{OR} \\ \bigwedge \Psi_{FT_T}(e') & \text{if } e \in \mathtt{IE} \text{ and } t(e) = \mathtt{AND} \\ \displaystyle\bigvee_{\substack{n_1,\ldots,n_k \\ n_1<\ldots<n_k}} \bigwedge_{i=1}^{k} \Psi_{FT_T}(e'_{n_i}) & \text{if } e \in \mathtt{IE} \text{ and } t(e) = \mathtt{VOT}(k/N) \end{cases}$$

*where* $\overline{\mathbf{B}}(v)$ *is a BDD with a single node in which* $Low(v) = \mathtt{0}$ *and* $High(v) = \mathtt{1}$.

## A.2 Algorithm 5: Translating FTs/formulae to BDDs

Following, the recursion scheme taken from [32] to translate FTs and layer one formulae is presented.

**Algorithm 5** Given $\phi$ and $T$, compute $\underline{\mathbf{B}}_T(\phi)$

**Input:** FT $T$, formula $\phi$
**Output:** $\underline{\mathbf{B}}_T(\phi)$
**Method:** Compute $\underline{\mathbf{B}}_T(\phi)$ according to the recursion scheme below. Store intermediate results $\underline{\mathbf{B}}_T(\cdots)$ and $\Psi_{FT_T}(\cdots)$ in a cache in case they are used several times.

**Recursion scheme:**

$\underline{\mathbf{B}}_T(e)$ : $\Psi_{FT_T}(e)$
$\underline{\mathbf{B}}_T(\neg\phi)$ : $\boldsymbol{\neg}(\underline{\mathbf{B}}_T(\phi))$
$\underline{\mathbf{B}}_T(\phi \wedge \phi')$ : $\underline{\mathbf{B}}_T(\phi) \boldsymbol{\wedge} \underline{\mathbf{B}}_T(\phi')$
$\underline{\mathbf{B}}_T(\phi[e_i \mapsto 0])$ : $\textsc{Restrict}(\underline{\mathbf{B}}_T(\phi), e_i, 0)$
$\underline{\mathbf{B}}_T(\phi[e_i \mapsto 1])$ : $\textsc{Restrict}(\underline{\mathbf{B}}_T(\phi), e_i, 1)$
$\underline{\mathbf{B}}_T(\mathrm{MCS}(\phi))$ : $\underline{\mathbf{B}}_T(\phi) \boldsymbol{\wedge} (\boldsymbol{\neg}\exists \mathtt{V}'.\mathtt{V}' \subset \mathtt{V} \boldsymbol{\wedge} \underline{\mathbf{B}}_T(\phi)[\mathtt{V} \curvearrowright \mathtt{V}'])$
$\underline{\mathbf{B}}_T(\mathrm{MPS}(\phi))$ : $\underline{\mathbf{B}}_T(\neg\phi) \boldsymbol{\wedge} (\boldsymbol{\neg}\exists \mathtt{V}'.\mathtt{V} \subset \mathtt{V}' \boldsymbol{\wedge} \underline{\mathbf{B}}_T(\neg\phi)[\mathtt{V} \curvearrowright \mathtt{V}'])$

$$\text{where: } \mathtt{V}' \subset \mathtt{V} \equiv \left(\bigwedge_k v'_k \Rightarrow v_k\right) \wedge \left(\bigvee_k v'_k \neq v_k\right)$$

where $\underline{\mathbf{B}}_T(\phi)[\mathtt{V} \curvearrowright \mathtt{V}']$ indicates the BDD $\underline{\mathbf{B}}_T(\phi)$ in which every variable $v_k \in \mathtt{V}$ is renamed to its primed $v'_k \in \mathtt{V}'$.

### A.3 Algorithm 6: Model checking PFL over a FT and a $\overline{b}$

**Overview.** As per [32], given a specific vector $\overline{b}$, a FT $T$ and a layer one formula $\phi$, this algorithm showcases how to check if $\overline{b}, T \models \phi$. To do so, we translate the given formula to a BDD and then we walk down the BDD from the root node following truth assignments given in the specific vector $\overline{b}$.

**Algorithm 6** Check if $\overline{b}, T \models \phi$, given $\overline{b}$, $T$ and $\phi$.

```
Input: boolean vector b̄ , FT T and a formula φ
Output: True iff b̄, T ⊨ φ, False otherwise.
Method: compute B_T(φ)
Starting from BDD root,
while current node w_i of B_T(φ) ∉ W_t do:
    if b_i ∈ b̄ = 0 then:
        w_i = Low(w_i)
    else if b_i ∈ b̄ = 1 then:
        w_i = High(w_i)
    end if
end while
if w_i = 0 then:
    return False
else if w_i = 1 then:
    return True
end if
```

**Algo. 6.** Algo. 6 shows an algorithm to check whether $\overline{b}, T \models \phi$, given a status vector $\overline{b}$, a FT $T$ and a formula $\phi$. A BDD for the formula $\phi$ is computed with regard to the structure function of the given FT $T$ i.e., we compute $\underline{\mathbf{B}}_T(\phi)$ as per Algo. 5. Subsequently, the algorithm walks down the BDD following the Boolean assignments given in $\overline{b}$: if the i-th element of $\overline{b}$ is set to 0 then the next node in the path will be given by $Low(w_i)$, if it is set to 1 then the next node will be $High(w_i)$. When the algorithm reaches a terminal node it returns *True* if its value is one - i.e., if $\overline{b}, T \models \phi$ - and *False* otherwise.

### A.4 Algorithm 7: Computing all satisfying vectors

**Overview.** Given a FT $T$ and a formula $\phi$, we now want to compute all vectors $\overline{b}$ such that $\overline{b}, T \models \phi$. In this scenario no Boolean vector is given. Thus, we need to construct the BDD for the given formula and then collect every path that leads to the terminal 1 to compute all satisfying vectors $[\![\overline{b}]\!]_T$ for the given formula.

**Algo. 7.** To achieve the desired outcome we will construct the BDD $\underline{\mathbf{B}}_T(\phi)$ for the given formula following Algo. 5. Then, the algorithm will walk down the BDD and store all the paths that lead to the terminal node 1. These paths represent all the status vectors that satisfy our formula $\phi$. The value for the elements of each vector is set to 0 or 1 if the stored path follows respectively the low or high edge of the collected elements of the BDD. After computing the BDD for a given $\phi$, AllSat [2] will achieve the desired outcome. This algorithm returns exactly all the satisfying assignments for a given BDD, i.e., in our case, all the Boolean vectors that satisfy our formula.

# B Appendix: Proofs

## B.1 Proof for Theorem 1

*Proof.* For a layer one formula $\phi$ and $\overline{\rho} \in B$, one can express

$$\Phi^*_T(\mu_{\overline{\rho}}, \phi) = \sum_{\substack{b \in \mathbb{B}^n: \\ \Phi_T(b,\phi)=1}} \prod_{i=1}^{n} \rho_i^{b_i}(1-\rho_i)^{1-b_i}. \tag{1}$$

This is a polynomial in the $n$ variables $\rho_i$. Each summand has degree 1 in each $\rho_i$, hence $\Phi^*_T(\mu_{\overline{\rho}}, \phi)$ can be written as

$$\Phi^*_T(\mu_{\overline{\rho}}, \phi) = \sum_{w \in \{0,1\}^n} c^h_w \prod_{i=1}^{n} \rho_i^{w_i} \tag{2}$$

for some constants $c^h_w \in \mathbb{R}$. Now fix an $i$, and let $\phi, \phi'$ be two Boolean formulae; then we can write $\frac{\Phi^*_T(\mu_{\overline{\rho}},\phi\wedge\phi')}{\Phi^*_T(\mu_{\overline{\rho}},\phi')} = \frac{A\rho_i+B}{C\rho_i+D}$ for some polynomials $A, B, C, D$ in the variables $\rho_1, \ldots, \rho_{i-1}, \rho_{i+1}, \ldots, \rho_n$. In particular, we have

$$\frac{\partial}{\partial \rho_i} \frac{\Phi^*_T(\mu_{\overline{\rho}}, \phi \wedge \phi')}{\Phi^*_T(\mu_{\overline{\rho}}, \phi')} = \frac{AD - BC}{(C\rho_i + D)^2}. \tag{3}$$

The sign of this partial derivative does not depend on the value of $\rho_i$. In particular, when all other $\rho_{i'}$ are fixed, this expression is maximized on an interval when $\rho_i$ is at one of the boundary points of that interval.

Now let us return to the setting of the Theorem; we will prove it for the maximum only as the minimum is proved analogously. Let Let $B = \prod_i [l_i, u_i]$ and let $\overline{\rho} \in \prod_i [l_i^{-1}, l_i^{+}]$; our aim is to find a vertex $\overline{\rho}'$ such that $\frac{\Phi^*_T(\mu_{\overline{\rho}},\phi\wedge\phi')}{\Phi^*_T(\mu_{\overline{\rho}},\phi')} \leq \frac{\Phi^*_T(\mu_{\overline{\rho}'},\phi\wedge\phi')}{\Phi^*_T(\mu_{\overline{\rho}'},\phi')}$. To do so, we construct a sequence $\overline{\rho}_0, \overline{\rho}_1, \ldots, \overline{\rho}_n$ with the following properties:

1. $\overline{\rho}_0 = \overline{\rho}$;
2. $\frac{\Phi^*_T(\mu_{\overline{\rho}_i},\phi\wedge\phi')}{\Phi^*_T(\mu_{\overline{\rho}_i},\phi')} \leq \frac{\Phi^*_T(\mu_{\overline{\rho}_{i+1}},\phi\wedge\phi')}{\Phi^*_T(\mu_{\overline{\rho}_{i+1}},\phi')}$ for $i < n$;
3. $\rho_{i,i'} \in \{l_{i'}, u_{i'}\}$ for $i' \leq i \leq n$.

This ensures that $\overline{\rho}' := \overline{\rho}_n$ has the required property. We define each $\overline{\rho}_i$ from $\overline{\rho}_{i-1}$ as follows: define $\overline{\rho}^-_i, \overline{\rho}^+_i \in [l_i, u_i]$ by

$$\overline{\rho}^{\bullet}_{i,i'} = \begin{cases} l_i, & \text{if } \bullet = - \text{ and } i' = i, \\ u_i, & \text{if } \bullet = + \text{ and } i' = i, \\ \rho_{i-1,i'}, & \text{if } i' \neq i. \end{cases}$$

By the discussion following (3), one has $\frac{\Phi^*_T(\mu_{\overline{\rho}_i},\phi\wedge\phi')}{\Phi^*_T(\mu_{\overline{\rho}_i},\phi')} \leq \max \left\{ \frac{\Phi^*_T(\mu_{\overline{\rho}^-_{i+1}},\phi\wedge\phi')}{\Phi^*_T(\mu_{\overline{\rho}^-_{i+1}},\phi')}, \frac{\Phi^*_T(\mu_{\overline{\rho}^+_{i+1}},\phi\wedge\phi')}{\Phi^*_T(\mu_{\overline{\rho}^+_{i+1}},\phi')} \right\}$.

Take $\overline{\rho}_{i+1} \in \{\overline{\rho}^-_{i+1}, \overline{\rho}^+_{i+1}\}$ to maximize $\frac{\Phi^*_T(\mu_{\overline{\rho}_{i+1}},\phi\wedge\phi')}{\Phi^*_T(\mu_{\overline{\rho}_{i+1}},\phi')}$, then this satisfies conditions 1–3 above. □